# Zipf's law, 1/$f$ noise, and fractal hierarchy

Yanguang Chen

(Department of Geography, College of Urban and Environmental Sciences, Peking University, 100871, Beijing, China. Email: chenyg@pku.edu.cn)

**Abstract:** Fractals, 1/$f$ noise, Zipf's law, and the occurrence of large catastrophic events are typical ubiquitous general empirical observations across the individual sciences which cannot be understood within the set of references developed within the specific scientific domains. All these observations are associated with scaling laws and have caused a broad research interest in the scientific circle. However, the inherent relationships between these scaling phenomena are still pending questions remaining to be researched. In this paper, theoretical derivation and mathematical experiments are employed to reveal the analogy between fractal patterns, 1/$f$ noise, and the Zipf distribution. First, the multifractal process follows the generalized Zipf's law empirically. Second, a 1/$f$ spectrum is identical in mathematical form to Zipf's law. Third, both 1/$f$ spectra and Zipf's law can be converted into a self-similar hierarchy. Fourth, fractals, 1/$f$ spectra, Zipf's law, and the occurrence of large catastrophic events can be described with similar exponential laws and power laws. The self-similar hierarchy is a more general framework or structure which can be used to encompass or unify different scaling phenomena and rules in both physical and social systems such as cities, rivers, earthquakes, fractals, 1/$f$ noise, and rank-size distributions. The mathematical laws on the hierarchical structure can provide us with a holistic perspective of looking at complexity such as self-organized criticality (SOC).

# 1. Introduction

Fractals, 1/$f$ noise (one-over-$f$ noise), Zipf's law, and the occurrence of large catastrophic events are typical ubiquitous general empirical observations across the individual sciences, which cannot be understood within the set of references developed within a specific scientific domain. We can find these scaling rules or phenomena here and there in physical and social systems, but it is hard to reveal the underlying rationale and universality of them. Bak (1996, page 12) was once puzzled by these observations: "Why are they universal, that is, why do they pop up everywhere? " These empirical observations have caused a broad research interest in the scientific circle. However, the inherent correlations between these scaling phenomena are still pending questions remaining to be explored despite the frequency with which them has been observed in nature and society.

In order to explain the ubiquity of the empirical observations associated with scaling laws, we must know the similarities and differences between fractals, $1/f$ spectra, Zipf's distribution, and the occurrence of large catastrophic events. Fractals suggest spatial patterns, $1/f$ noise implies temporal processes, Zipf's law indicates hierarchical structure, and the occurrence of large catastrophic events may merge into the Zipf distribution because it can be described with Zipf's law. The abovementioned patterns, processes, structure and distribution share similar scaling laws, maybe they can be mathematically converted into one another. For example, a fractal pattern can be transformed into a hierarchical process by recursive subdivision of space (Batty and Longley, 1994; Chen, 2008; Goodchild and Mark, 1987). If the fractal is a multi-scaling pattern, the result of spatial recursive subdivision may follow Zipf's law empirically. If so, we will be able to construct the theoretical relations between different scaling rules or phenomena including fractals, Zipf's law, and $1/f$ fluctuation.

Recently years, there are many interesting researches about Zipf's law, $1/f$ noise, and fractals (e.g. Altmann *et al*, 2009; Alvarez-Martinez et al, 2010;Amaral *et al*, 2004; Batty, 2005; Batty, 2006; Batty, 2008; Blasius and Tönjes, 2009; Chen, 2008; Eliazar and Klafter, 2009a; Eliazar and Klafter, 2009b; Ferrer i Cancho and Solé, 2003; Ferrer i Cancho *et al*, 2005; Jiang, 2009; Jiang and Jia, 2011), but little study tries to combine or unify them. This paper will reveal the analogy between fractals, $1/f$ noise, and Zipf's law, and resolve the following problems. First, Zipf's law can be employed to estimate the capacity dimension of multifractals. Second, both the $1/f$ spectra and Zipf distribution will be transformed into a hierarchy with cascade structure. Third, a new framework is proposed to understanding the simple rules of complex systems. This study will provide scientists with an integrated framework of looking at how complex systems are organized to work, especially the self-organized criticality (SOC) (Bak, 1996).

## 2. The self-similar hierarchy of Cantor ternary set

### 2.1 Mono-fractal structure

A regular fractal is a typical hierarchy with cascade structure. Let's take the well-known Cantor ternary set as an example to show how to describe the hierarchical structure and how to calculate its fractal parameter. The Cantor set, sometimes also called the Cantor comb or no middle third set (Cullen, 1968), can be created by repeatedly deleting the open middle thirds of a set of line segments (Mandelbrot, 1983; Peitgen *et al*, 2004). In theory, the Cantor set comprises all points in the closed interval [0, 1] which are not deleted at any step in the infinite process (Figure 1). Formally, we can use two measurements, the length ($L$) and number ($N$) of fractal copies in the $m$th level, to characterize the self-similar hierarchy. Thus, we have two exponential functions such as

$$f_m = N_1 r_n^{m-1} = \frac{N_1}{r_n} e^{(\ln r_n)m} = N_0 e^{\omega m}, \tag{1}$$

$$L_m = L_1 r_l^{1-m} = L_1 r_l e^{-(\ln r_l)m} = L_0 e^{-\psi m}, \tag{2}$$

where $m$ denotes the ordinal number of class ($m$=1, 2, …), $N_m$ is the number of the fractal copies of a given length, $L_m$ is the length of the fractal copies in the $m$th class, $N_1$ and $L_1$ are the number and length of the initiator ($N_1$=1), respectively, $r_n$ and $r_l$ are the **number ratio** and **length ratio** of fractal copies, $N_0=N_1/r_n$, $L_0=L_1 r_l$, $\omega=\ln(r_n)$, $\psi=\ln(r_l)$. Apparently, the common ratios are

$$r_n = \frac{N_{m+1}}{N_m} = \frac{N_1 r_n^m}{N_1 r_n^{m-1}} = 2, \quad r_l = \frac{L_m}{L_{m+1}} = \frac{L_1 r_l^{1-m}}{L_1 r_l^{-m}} = 3.$$

From equations (1) and (2) follows

$$N_m = \mu L_m^{-D}, \tag{3}$$

in which $\mu=N_1 L_1^D$ is the proportionality coefficient, and $D$ is the fractal dimension of the Cantor set. Now, three approaches to estimating the fractal dimension can be summarized as follows. Based on the power law, the fractal dimension can be expressed as

$$D = -\frac{\ln(N_{m+1}/N_m)}{\ln(L_{m+1}/L_m)}, \tag{4}$$

Based on the exponential models, the fractal dimension is

$$D = \frac{\omega}{\psi}. \tag{5}$$

Based on the common ratios, the fractal dimension is

$$D = \frac{\ln r_n}{\ln r_l}. \tag{6}$$

In light of both the mathematical derivation and the empirical analysis, we have

$$D = -\frac{\ln(N_{m+1}/N_m)}{\ln(L_{m+1}/L_m)} = \frac{\ln r_n}{\ln r_l} = \frac{\omega}{\psi} = \frac{\ln(2)}{\ln(3)} \approx 0.6309.$$

This suggest that, for the regular fractal hierarchy, the fractal dimension can be computed by using exponential functions, power function, or common ratios, and all these results are equal to one another.

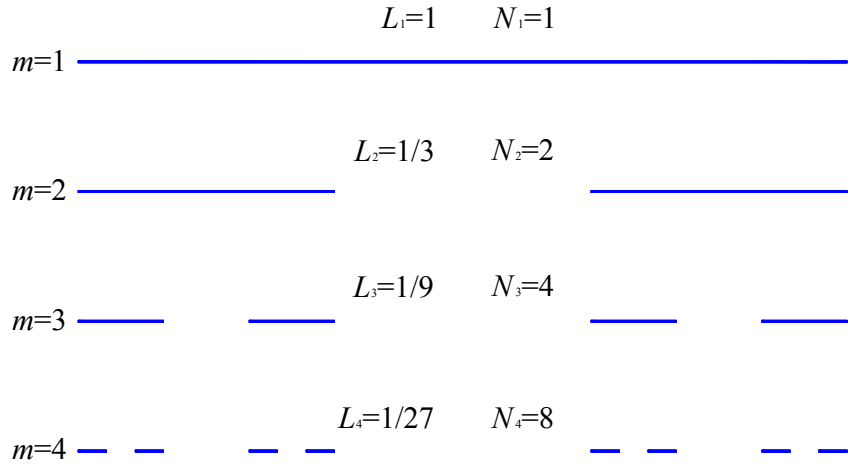

**Figure 1** The Cantor set as a self-similar hierarchy with cascade structure (the first four classes)

If we use length to measure the size of fractal copies, the simple self-similar hierarchy of Cantor set will give a hierarchical step-like frequency distribution rather than the continuous frequency curve suggested by the rank-size rule (Figure 2). In other words, the mono-fractal hierarchy of Cantor set fails to follow Zipf's law. However, if we substitute the multifractal structure for mono-fractal structure, the multi-scaling Cantor set will empirically follow the rank-size rule in a broad sense.

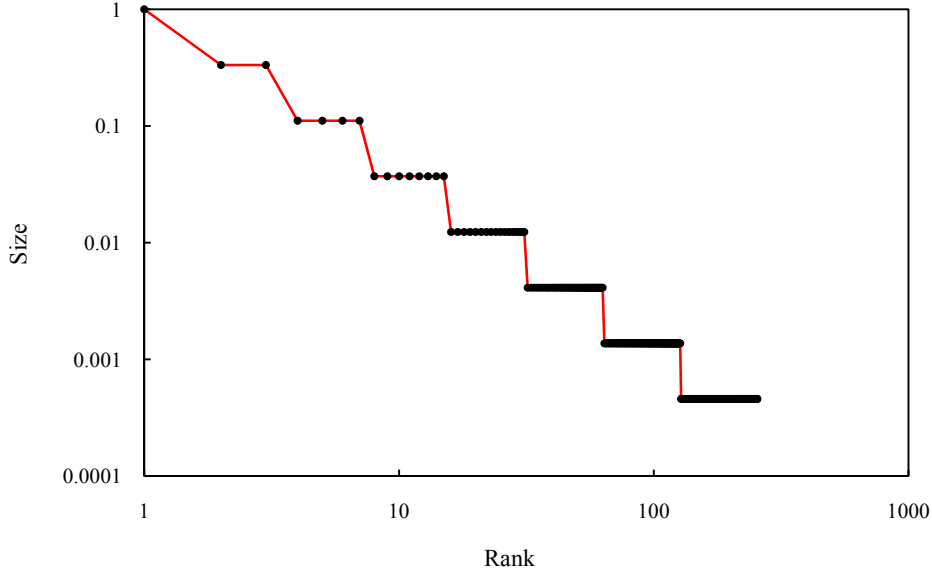

**Figure 2** The step-like rank-size distribution of the mono-fractal Cantor set (255 data points)

## 2.2 Multi-fractal structure

It is not difficult to construct a multifractal hierarchy based on the Cantor set. In the top class, the initiator is a straight line segment in unit length, that is, $S_1=L_1=1$. In the second class, the generator includes two straight line segments with different lengths. One segment length is $a$, and the other, $b$. let $a=0.375$, $b=2/3-a$. The length summation of the two line segments is 2/3, namely, $S_2=a+b=2/3$, and the average length is, $L_2= S_2/2=1/3$. In the third class, there are four line segments, and the lengths are $a^2$, $ab$, $ba$, and $b^2$, respectively. The total length of these line segments is 4/9, namely, $S_3=(a+b)^2=(2/3)^2$, and the average length is $L_2= S_2/4=1/3^2$ (Figure 3). Generally speaking, the $m$th class consists of $2^{m-1}$ line segments with lengths of $a^{m-1}$, $a^{m-2}b$, $a^{m-3}b^2$, ..., $a^2b^{m-3}$, $ab^{m-2}$, and $b^{m-1}$, respectively (Table 1). The length summation is

$$S_m = (a+b)^{m-1} = (\frac{2}{3})^{m-1}; \qquad (7)$$

The average length is

$$L_m = \frac{(a+b)^{m-1}}{N_m} = 3^{m-1}. \qquad (8)$$

Combining equations (7) and (8) yields such a scaling relation

$$N_m = L_m^{-\ln(2)/\ln(3)} = \mu L_m^{-D_0}, \tag{9}$$

which is identical in form to equation (3), and the capacity dimension $D_0=\ln(2)/\ln(3)\approx 0.631$ is equal to the fractal dimension of the simple Cantor set displayed in Figure 1.

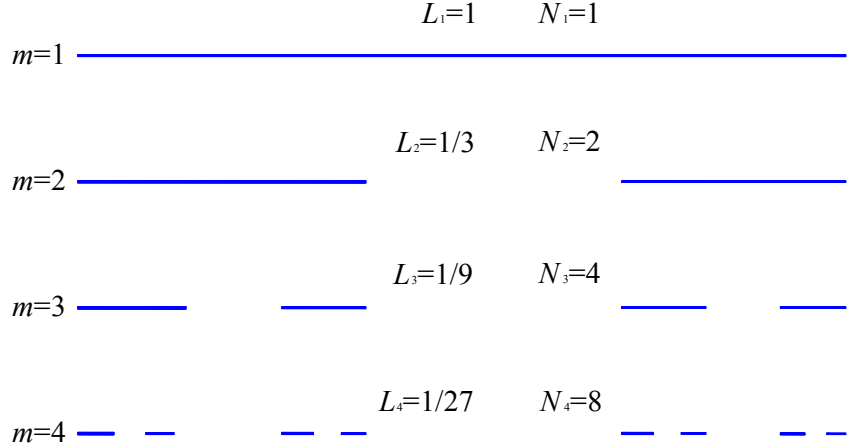

**Figure 3** The two-scale Cantor set as a complex hierarchy with cascade structure (The first four classes are shown to indicate the first four steps)

The multifractal Cantor set is a complex hierarchy differing from the simple mono-fractal Cantor set. We can demonstrate that the length of the fractal copies in the complex Cantor set follow general Zipf's law, a rank-size scaling law. The first 11 classes contains 2047 fractal copies, and scaling relation of rank and size (length) is as follows

$$L_k = L_1 k^{-v} = 1.866 k^{-1.598}. \tag{10}$$

where $k$ is rank of fractal copies, $L_k$ is the length of the $k$th copies, $L_1$ is the proportionality coefficient, and $v$, the scaling exponent. The exponent value $1/v \approx 1/1.598 \approx 0.626 \to D \approx 0.631$, and the goodness of fit is about $R^2=0.996$. The more the number of fractal copies is, the closer the scaling exponent is to the reciprocal of the fractal dimension of Cantor set. The scaling pattern is very similar to the rank-size pattern of cities except that the scaling exponent is not close to 1 (Figure 4). In fact, if we only consider the first 1800 data points within the scaling range, the scaling exponent is about $v \approx 1/1.581$, thus the reciprocal $1/v \approx 0.633$ is closer to the capacity dimension value. This suggests that the scaling range is very important for us to estimate fractal dimension properly.

**Table 1** The full length, average length, and numbers of fractal copies in different steps of Cantor set

| Level $m$ | Full length $S_m$ | Average length $L_m$ | Number $f_m$ |
|---|---|---|---|
| 0 | $(2/3)^0$ | $(1/3)^0$ | $2^0=1$ |
| 1 | $(2/3)^1$ | $(1/3)^1$ | $2^1=2$ |
| 2 | $(2/3)^2$ | $(1/3)^2$ | $2^2=4$ |
| 3 | $(2/3)^3$ | $(1/3)^3$ | $2^3=8$ |
| 4 | $(2/3)^4$ | $(1/3)^4$ | $2^4=16$ |
| 5 | $(2/3)^5$ | $(1/3)^5$ | $2^5=32$ |
| 6 | $(2/3)^6$ | $(1/3)^6$ | $2^6=64$ |
| 7 | $(2/3)^7$ | $(1/3)^7$ | $2^7=128$ |
| 8 | $(2/3)^8$ | $(1/3)^8$ | $2^8=256$ |
| 9 | $(2/3)^9$ | $(1/3)^9$ | $2^9=512$ |
| … | … | … | … |

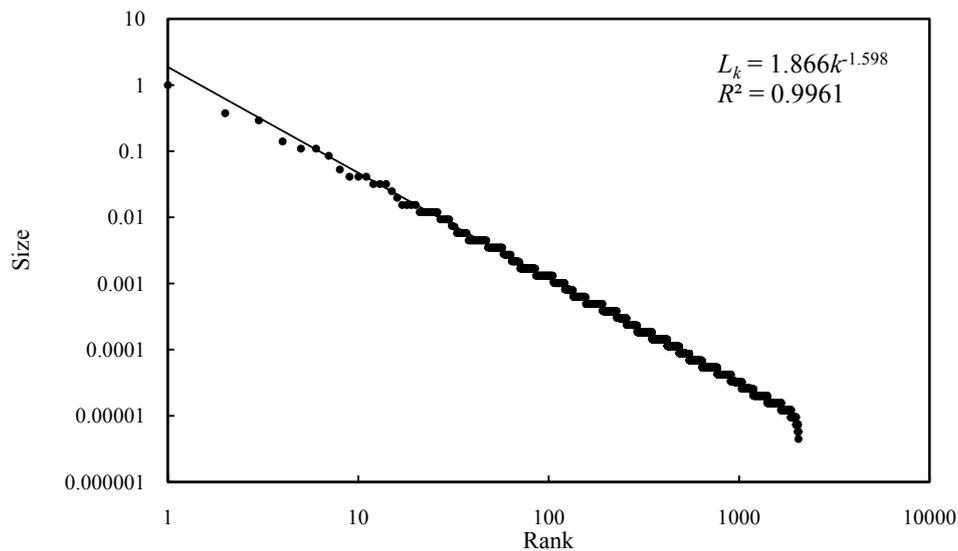

**Figure 4** The rank-size distribution of multi-fractal Cantor set (511 data points)

In practice, the capacity dimension of the multi-scale Cantor set can be estimated with the Pareto distribution. For example, for the 2047 fractal copies based on 11 levels, on the largest 1408 data points can be properly measured if the scale is taken as $L_m=(1/3)^{m-1}$. Consequently, by the least square computation, the number of the fractal copies greater than $L_m$ is formulated as $N(L)=1.043L^{-0.667}$, which gives the Pareto exponent approximation 0.667, and the goodness of fit is about $R^2=0.998$. The more the data points are, the closer the Pareto scaling exponent is to the fractal dimension $D_0=0.631$. This suggests that the Pareto distribution is an alternative approach of Zipf's law to estimating the capacity dimension of complex Cantor set.

## 3. The geometric subdivision principle of the harmonic sequence

### 3.1 Zipf's law, 1/$f$ spectra, and the harmonic sequence

Zipf's law is originally proposed as empirical formulation describing the frequency-rank scaling relation in linguistics (Zipf, 1949), but it actually stemmed from the rank-size rule of earlier Auerbach's urban studies in 1913 (Carroll, 1982). Gabaix (1999) once pointed out: "Zipf's law for cities is one of the most conspicuous empirical facts in economics, or in the social sciences generally." If the cities in a region are rank-ordered in terms of their populations from the largest (1) to the smallest ($N$), then the population of the city ranked $k$ can be derived from

$$P_k = P_1 k^{-q}, \qquad (11)$$

where $P_k$ refers to the population of the city ranked $k$ and $P_1$ to the population of the largest city ($k$=1,2,3,…,$N$), $q$ is a scaling exponent usually close to 1 (Gabaix, 1999). Nowadays, Zipf's law has become a well-known empirical law, suggesting the fact that many types of data studied in various sciences can be approximated with a rank-size scaling (Bettencourt *et al*, 2007). Zipf's law proved to be a fundamental law in natural languages (Ferrer-i-Cancho and Elvevåg, 2010), from which follows other scaling laws such as Heaps' law (Lü *et al*, 2010; Serrano *et al*, 2009).

The Zipf distribution is one of a family of related power-law probability distributions in physical and social systems. Another important discrete power law is the frequency-spectrum scaling relation of random processes or signals. The general form of the frequency-spectrum scaling is

$$S(f) = S_0 f^{-\beta}, \qquad (12)$$

in which $f$ refers to frequency, $S(f)$ to the power spectral density, and $\beta$ to the spectral exponent, and $S_0$ is proportionality coefficient. The value of the spectral exponent comes between 0 and 2. When $\beta$=0, we have a white noise (1/$f^0$ noise); when $\beta$=2, we have a red noise or Brownian noise (1/$f^2$ noise). In the most cases, $\beta$ approaches 1 and we have a pink noise (1/$f^1$ noise) (Bak, 1996; Mandelbrot, 1999). If the length of the sample path from a random process is $T$, the discrete frequency can be expressed as $f=k/T$ ($k$=1, 2, 3, …, $T$/2). Thus equation (12) can be empirically

rewritten as

$$S_k = S_0(\frac{k}{T/2})^{-\beta} = S_1 k^{-\beta}, \quad (13)$$

where $S_1=S_0(T/2)^{\beta}$, and $S_k$ denotes the discrete result of $S(f)$.

For the random process, the spectral density is always inversely proportional to its frequency; for the rank-size distribution, the size is always inversely proportional to its rank. In short, both the rank-size scaling exponent $q$ and the spectral exponent $\beta$ are always close to 1, that is, $q\rightarrow 1$, $\beta\rightarrow 1$. If $q=1$, Zipf's law suggests a rank-size rule or a $1/k$ distribution (Carroll, 1982; Knox and Marston, 2006); if $\beta=1$, we have a $1/f$ noise or a 1-over-$f$-like spectrum (Bak, 1996; Bak *et al*, 1987; Chen, 2008). For simplicity, let $P_1=1$, $S_1=1$, $q=1$, and $\beta=1$. Thus both the size distribution and spectral density can be abstracted as a harmonic sequences, $\{1/k\}$, where $k=1, 2, 3,\ldots$. Then we can divide the harmonic sequences into $M$ classes in a top-down way according to the $2^n$ rule (Davis, 1978; Chen and Zhou, 2004). The result is one for the first class, two fractions for the second class, four fractions for the third class, and so on (Figure 5). This subdivision will yield a hierarchy with cascade structure, and the first four classes is as below:

$$m = 1: \quad 1$$
$$m = 2: \quad \frac{1}{2}, \frac{1}{3}$$
$$m = 3: \quad \frac{1}{4}, \frac{1}{5}, \frac{1}{6}, \frac{1}{7}$$
$$m = 4: \quad \frac{1}{8}, \frac{1}{9}, \frac{1}{10}, \frac{1}{11}, \frac{1}{12}, \frac{1}{13}, \frac{1}{14}, \frac{1}{15}$$
……

Generally, the $m$th class is $[1/2^{m-1}, 1/(2^{m-1}+1),\ldots,1/(2^m-1)]$, where $m=1, 2, \ldots, M$, $M$ goes to infinity in theory. Based on the self-similar hierarchy, the rank-size scaling and frequency-spectrum scaling can be reformulated. According to the postulate of the $2^n$ rule ($n=1, 2, 3, \ldots$), the number of elements of order $m$, $N_m$, is defined as follows

$$N_m = N_1 r_n^{m-1} = 2^{m-1}, \quad (14)$$

where $N_1=1$ denotes the number in the top class. Obviously, the interclass *number ratio* is

$$r_f = \frac{N_{m+1}}{N_m} = 2. \quad (15)$$

If we can prove that the sum of numbers in each class ($S_m$) is a constant, that is $S_m=N_m P_m=const$,

then we will have an average value $P_m=const/2^{m-1}$. This suggests that the rank-size rule as well as the frequency-spectrum scaling is mathematically equivalent to the $2^n$ rule.

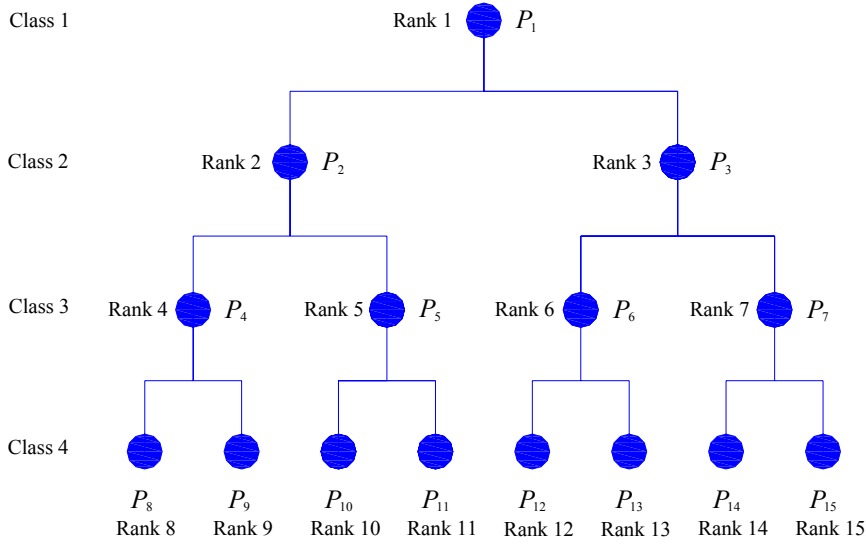

**Figure 5** A schematic diagram of the self-similar hierarchy of cities based on the $2^n$ principle (The first four classes, see Chen and Zhou, 2004, or Jiang and Yao, 2010)

### 3.2 A derivation of the $2^n$ rule from the harmonic sequence

The derivation of the $2^n$ rule from the harmonic sequence can be abstracted as mathematical proof problem. According to the geometric regulation of subdividing the harmonic sequence into different levels, the sum of the fractions in the $m$th class, $S_m$, can be expressed as

$$S_m = \frac{1}{2^{m-1}} + \frac{1}{2^{m-1}+1} + \cdots + \frac{1}{2^m-1} = \sum_{j=0}^{2^{m-1}-1} \frac{1}{2^{m-1}+j}, \qquad (16)$$

where $j=0, 1, 2, \ldots, 2^{m-1}-1$. By mathematical experimentation, we can find that if $m=M$ is large enough, the summation in each level approaches $\ln(2) \approx 0.6931$. The derivation of the $2^n$ rule will be complete if we can prove

$$\lim_{m \to \infty} S_m = \ln 2. \qquad (17)$$

The mathematical proof is very simple. In fact, equation (16) can be rewritten as

$$S_m = \frac{1}{2^{m-1}(1+\frac{0}{2^{m+1}})} + \frac{1}{2^{m-1}(1+\frac{1}{2^{m+1}})} + \cdots + \frac{1}{2^{m-1}(1+\frac{2^{m-1}-1}{2^{m-1}})} = \sum_{j=0}^{2^{m-1}-1} \frac{\frac{1}{2^{m-1}}}{1+\frac{j}{2^{m-1}}}. \qquad (18)$$

Let $f(x)=1/(1+x)$, which is defined in the closed interval [0, 1]. Then we can use the fractions

$$0 = \frac{0}{2^{m-1}}, \frac{1}{2^{m-1}}, \frac{2}{2^{m-1}}, \cdots, \frac{2^{m-1}-1}{2^{m-1}}, \frac{2^{m-1}}{2^{m-1}} = 1, \quad (19)$$

to divide the interval [0, 1] into $2^{m-1}$ equal parts with a length of $1/2^{m-1}$. Taking

$$x_j = \frac{j}{2^{m-1}} \quad (20)$$

suggests

$$\Delta x_j = \frac{1}{2^{m-1}}. \quad (21)$$

Further, if $m$ is big enough, we will have

$$x_j = \lim_{m \to \infty} \frac{2^{m-1}-1}{2^{m-1}} = 1, \quad (22)$$

which implies $x_j \in [0,1]$. Under the condition of limit, the summation of order $m$ is

$$\lim_{m \to \infty} S_m = \lim_{m \to \infty} \sum_{j=0}^{2^{m-1}-1} \frac{\frac{1}{2^{m-1}}}{1+\frac{j}{2^{m-1}}} = \lim_{m \to \infty} \sum_{j=0}^{2^{m-1}-1} \frac{\Delta x_j}{1+x_j} = \int_0^1 \frac{1}{1+x} dx = [\ln(1+x)]_0^1 = \ln 2. \quad (23)$$

Thus the proof is ended and the conclusion can be obtained.

In terms of equation (23), we have

$$S_m = N_m A_m = \ln 2. \quad (24)$$

where $N_m$ is the number of elements and $A_m$ the average value of the fractions in the $m$th level. Considering equation (14), the average value in each class is

$$A_m = \frac{S_m}{N_m} = \ln 2 / 2^{m-1} = P_1 r_a^{1-m}. \quad (25)$$

Apparently, the interclass *size ratio* is

$$r_a = \lim_{m \to \infty} \frac{A_m}{A_{m+1}} = \frac{N_{m+1}}{N_m} = 2. \quad (26)$$

This implies that the size ratio $r_a$ equals the number ratio $r_f$. From equations (14) and (25) follows

$$N_m = \eta A_m^{-D}, \quad (27)$$

in which $\eta = N_1 A_1^D = 1$ refers to proportionality coefficient, and $D=1/q$ or $D=1/\beta$ to the fractal dimension of the hierarchy, which can be formulated as

$$D = \frac{\ln r_n}{\ln r_a} \to 1. \tag{28}$$

where "→" denotes "be close to".

In theory, if $m=1$ and $D=1$, then we have $\eta=N_1A_1=1>\ln(2)$. However, comparing equation (27) with equation (24) shows $\eta=S_m=\ln(2)$ instead of 1. This suggests that only when $m$ is large enough, the rank-size rule and the frequency-spectrum scaling rule is equivalent to the $2^n$ rule. In other words, if we transform a rank-size distribution or a frequency-spectrum distribution into a self-similar hierarchy, the first several classes always depart from the scaling range to some extent theoretically. As for the empirical data, the last class always gets out of the scaling range because of undergrowth of small elements. Therefore, the exponential laws as well as the power laws of hierarchies are usually invalid at the extreme scales, i.e. the very large and small scales.

## 4. Mathematical experiments

### 4.1 The mathematical experiment based on Zipf's law

The equivalence relation between the $2^n$ rule and the rank-size rule or the frequency-spectrum scaling rule can be easily testified by simple mathematical experiments through computer software, say, MS Excel. The summation of the first class is 1, second class, 1/2+1/3=0.8333, the third class, 1/4+1/5+1/6+1/7=0.7595, and so forth. If $m$ is large enough, we will have total number in $m$th class $N_mA_m=\ln(2)\approx 0.6931$ in theory. By limit analysis, the size ratio is $r_a=r_n=A_m/A_{m+1}=N_{m+1}/N_m=2$. Accordingly, the fractal dimension of the hierarchy with cascade structure is $D=\ln r_n/\ln r_a \to 1$. The first 10 classes of a mathematical experiment result is listed in Table 2 and displayed in Figure 6.

Table 2 The partial result of mathematical experiment by converting the harmonic sequences based on the rank-size rule into geometric sequences based on the $2^n$ rule ($r_n=2$)

| Level ($m$) | Element number ($N_m$) | Size sum ($N_mA_m$) | Average size ($A_m$) | Size ratio ($r_a$) | Standardized size ratio ($r_a^*$) |
|---|---|---|---|---|---|
| 1 | 1 | 1 | 1 | | |
| 2 | 2 | 0.8333 | 0.4167 | 2.400 | 2.352 |
| 3 | 4 | 0.7595 | 0.1899 | 2.194 | 0.809 |
| 4 | 8 | 0.7254 | 0.0907 | 2.094 | 0.057 |
| 5 | 16 | 0.7090 | 0.0443 | 2.046 | -0.303 |
| 6 | 32 | 0.7010 | 0.0219 | 2.023 | -0.478 |
| 7 | 64 | 0.6971 | 0.0109 | 2.011 | -0.564 |
| 8 | 128 | 0.6951 | 0.0054 | 2.006 | -0.607 |
| 9 | 256 | 0.6941 | 0.0027 | 2.003 | -0.628 |
| 10 | 512 | 0.6936 | 0.0014 | 2.001 | -0.639 |
| … | … | … | … | … | … |
| $M$ | $2^{M-1}$ | $\ln(2)$ | $\ln(2)/(2^{M-1})$ | 2.000 | $-\ln(2)$ |

**Note**: According to standardized data of the size ratios, the first and second class can be treated as an outlier since the $r_a^*$ value is greater than 2. This is consistent with the conclusion from mathematical proof.

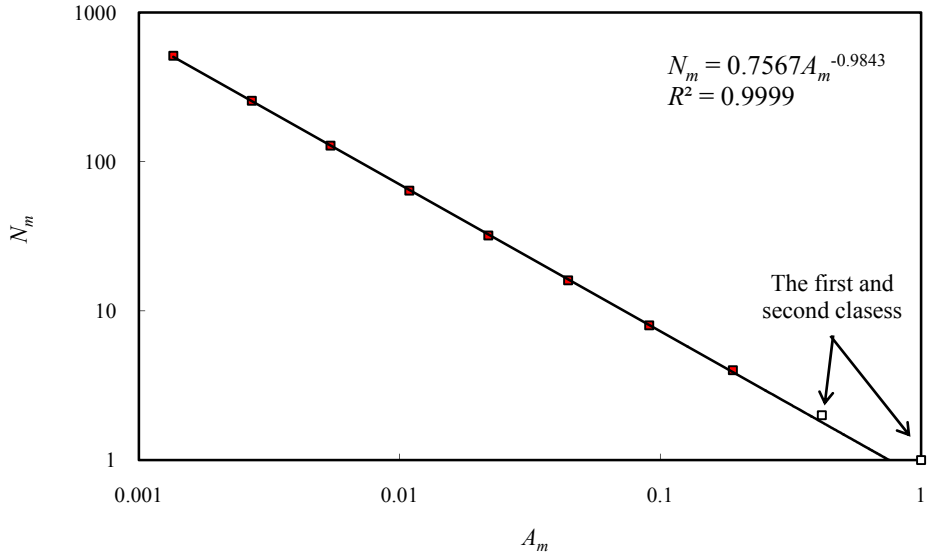

**Figure 6** The scaling relation between element numbers and average sizes of different classes (Only the first 10 classes is considered in the $2^n$ hierarchy)

By mathematical experiments, we can make two generalizations. First, if the rank-size scaling exponent $q=1$ or the frequency-spectral exponent $\beta=1$, the $2^n$ rule can be easily generalized to the $3^n$ rule, the $4^n$ rule, the $5^n$ rule, and the $N^n$ rule ($n=1, 2, 3, …$; $N=2, 3, 4, …$). That is, if the number ratio $r_n=2$ is replaced by $r_n=N$, where $N$ is an integer greater than 1, all the above-mentioned relations and results are still true, and we will have a self-similar hierarchy with a size ratio $r_a \rightarrow N$. Second, if $q \neq 1$ or $\beta \neq 1$, the rank-size distribution or the frequency-spectrum distribution can still be transformed into a self-similar hierarchy, and we have $r_n=2$, $r_a=2^q$ or $r_a=2^\beta$. In this case, we have a generalized $2^n$ rule which cannot be generalized to the $N^n$ rule ($n=1, 2, 3, …$; $N=3, 4, 5, …$).

Further, we can perform mathematical experiments to verify the $3^n$ rule, the $4^n$ rule, and generally, the $N^n$ rule for the scaling exponent equal to 1. We can also verify the generalized $2^n$ rule for the arbitrary scaling exponent ($0<q<3$, $0<\beta<3$).

## 4.2 The mathematical experiment for the 1/*f* spectra

For the standard form, a 1/*f* noise is mathematically equivalent to a 1/*k* distribution. In order to yield a regular 1/*f* spectrum, we may as well replace the 1/*f* noise by an inverse power function in the form

$$x_t = x_1 t^{-\alpha}, \tag{29}$$

where $t$ denotes time, $x_t$ is some kind of density value at time $t$, $x_1$ and $\alpha$ are parameters. Let $x_1=100$, $\alpha=0.3$, and $t=1, 2, \ldots, 1024$, the spectral density follow the inverse power law such as

$$\hat{S}(f) = 1.660 f^{-1.039}. \tag{30}$$

The goodness of fit is about $R^2=0.994$ (Figure 7). This is what is called 1/*f* spectrum. The frequency-spectrum relation can be equivalently converted to a rank-size relation in the form

$$\hat{S}_k = 2220.542 k^{-1.039}. \tag{31}$$

The correlation coefficient square has no change. It is easy to change the frequency-spectrum scaling into the size-number scaling (Table 3). Thus, equation (30) can be rewritten as

$$\hat{N}_m = 134.31 A_m^{-1.008}. \tag{32}$$

The goodness of fit is about $R^2=0.996$ (Figure 8).

**Table 3** The partial result of mathematical experiment by converting the spectral density based on an inverse power law into geometric sequences based on the $2^n$ rule ($r_n=2$)

| Level ($m$) | Element number ($N_m$) | Size sum ($N_m A_m$) | Average size ($A_m$) | Size ratio ($r_a$) |
|---|---|---|---|---|
| 1 | 1 | 110.5420 | 110.5420 | |
| 2 | 2 | 121.4519 | 60.7260 | 1.8203 |
| 3 | 4 | 134.2228 | 33.5557 | 1.8097 |
| 4 | 8 | 145.7016 | 18.2127 | 1.8424 |
| 5 | 16 | 152.1348 | 9.5084 | 1.9154 |
| 6 | 32 | 150.1415 | 4.6919 | 2.0266 |
| 7 | 64 | 138.6331 | 2.1661 | 2.1660 |
| 8 | 128 | 122.2321 | 0.9549 | 2.2684 |
| 9 | 256 | 120.6764 | 0.4714 | 2.0258 |

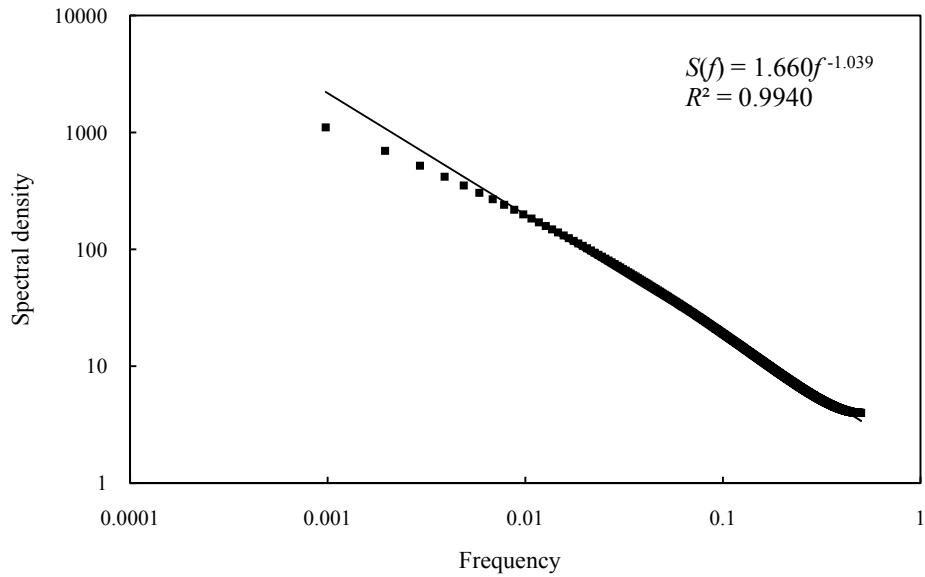

**Figure 7** The frequency-spectrum scaling relation based on an inverse power law distribution ($t$ ranges from 1 to 1024, $f$ varies from 1/1024 to 1/2)

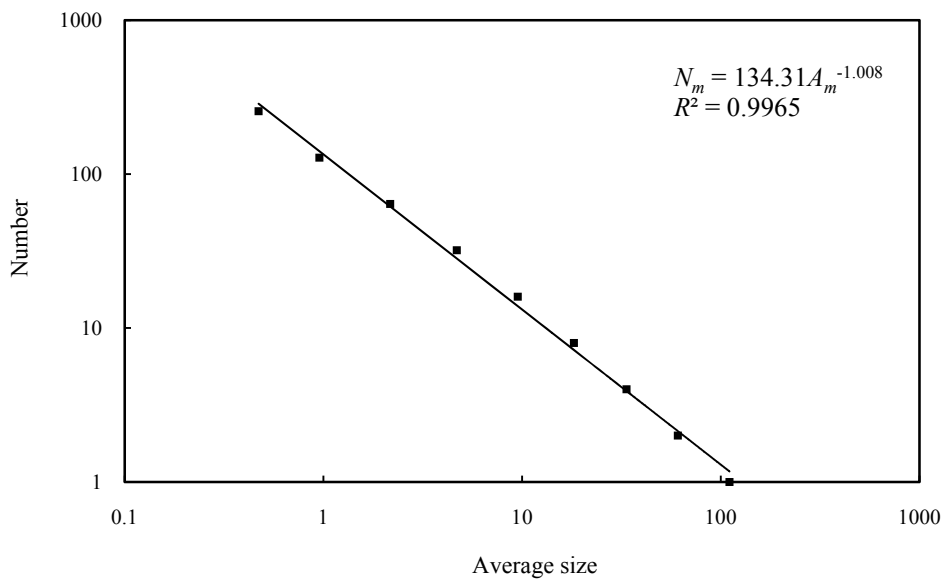

**Figure 8** The size-number scaling relation based on the $1/f$ distribution ($f$ varies from 1/1024 to 511/1024)

## 5. Discussion and conclusions

Now we can see three kinds of typical hierarchies. The first is mono-fractal hierarchy (e.g. Figure 1), the second is multifractal hierarchy (e.g. Figure 3), and the third, rank-size hierarchy (e.g. Figure 5). The common characteristics of these hierarchies are cascade structure and scaling invariance. The first one is a pure mathematical pattern which cannot be found in the real world. The second and third one can be employed to model the real scale-free phenomena in natural and human systems. In practice, it is hard to distinguish between the multifractal hierarchy and the rank-size hierarchy. Using the same method as generating the multi-scaling Cantor set, we can generate a multifractal rank-size distribution by revising equation (7) as

$$S_m = (a+b)^{m-1} = 1. \tag{33}$$

The result is very similar to the real rank-size distribution of cities (Figure 9). The size distribution of cities can be modeled by either the rank-size hierarchy or the multifractal hierarchy (Chen and Zhou, 2003; Chen and Zhou, 2004). Where city development or urban evolution is concerned, there seems to be a competition process, and the rank-size distribution of cities seems to struggle between the multifractal hierarchy and the rank-size hierarchy.

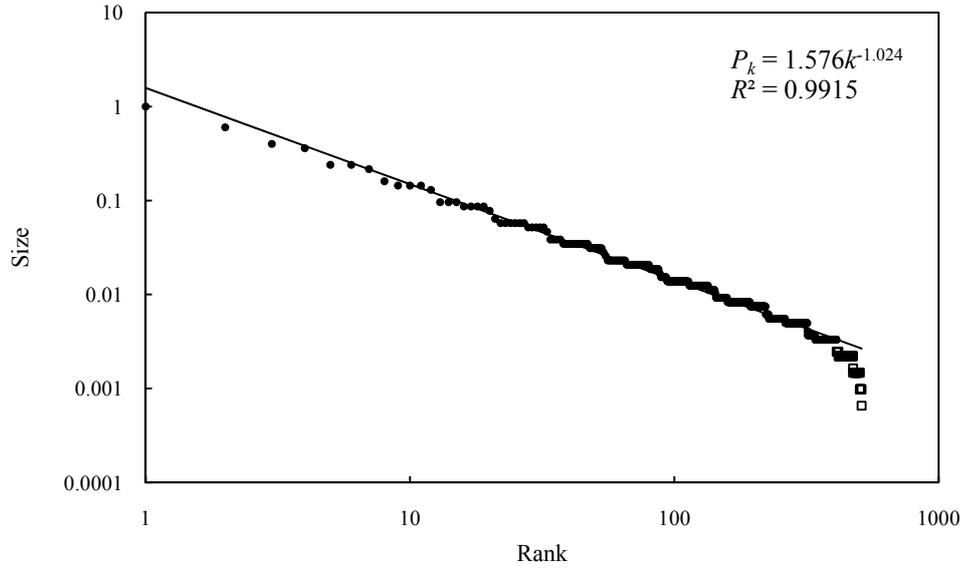

**Figure 9** The multifractal rank-size distribution of cities for *a*=0.6 and *b*=0.4 (The system consists of 511 elements, but only the largest 410 ones are within the scaling range. The solid points indicate the scale-free range)

The self-similar hierarchy is a general framework which is associated with various scaling rules and phenomena. The hierarchy always follows a pair of exponential laws and a power law. Actually, equations (14) and (25) can be generalized to more universal forms such as

$$y_m = y_1 r_n^{m-1}, \tag{34}$$

$$x_m = x_1 r_a^{1-m}, \tag{35}$$

where *x* refers to some size measurement, and *y* to some number measurement, $r_a$, $r_n$ are common ratios ($r_a>1$, $r_n>1$), and $x_1$, $y_1$ are parameters. If we use equations (34) and (35) to describe river systems, we have Horton-Strahler's laws (Horton, 1945; Strahler, 1952); if we use equations (34) and (35) to characterize the earthquake energy distribution within given region and period, we have Gutenberg-Richter's laws (Gutenberg and Richter, 1954); if we use equations (34) and (35) to model cities in a country, we have Davis' $2^n$ rule and the generalized $2^n$ rule (Chen and Zhou, 2003; Davis, 1978). From equations (34) and (35) follows a general power law

$$y_m = \eta x_m^{-D}, \qquad (36)$$

which can be used to reflect fractal structure of scale-free systems. Thus the self-similar hierarchy can be employed to encompass fractals, $1/f$ noise, Zipf's law, and the occurrence of large catastrophic events (e.g. earthquakes). Equations (34) and (35) can be derived from the principle of entropy-maximizing (Carroll, 1982; Chen, 2008; Curry, 1964; Wilson, 2000). This suggests that the physical fundament of equations (36) rests with entropy maximization. The principle of entropy maximizing is possibly associated with the principle of least action, i.e., the principle of least effort (Ferrer i Cancho and Solé, 2003; Zipf, 1949), but this remains to be further explored in future.

The main points of this paper can summarized as follows. First, fractal process can be treated as a hierarchy with cascade structure, but only the fractal copies in a multifractal system meet the rank-size distribution formally, and we can use the Pareto distribution or Zipf's law to estimate the capacity dimension of the multifractals. Second, both the Zipf distribution and $1/f$ spectra can be converted into self-similar hierarchies. By the hierarchical structure, we can find more than one approach to estimating the scaling exponents, and associate the scaling laws with the principle of entropy-maximization. Third, the self-similar hierarchies fall into three types—the mono-fractal hierarchy with step-like rank-size distribution, the multifractal hierarchy with quasi-continuous or semi-continuous rank-size distribution, and the rank-size hierarchy consisting of "harmonic sequence". The last one is also one of the mono-fractal phenomena. Fourth, the self-similar hierarchy is a more general pattern which can be employed to unify the ubiquitous empirical observations such as fractal patterns, $1/f$ spectra, Zipf's law, and the occurrence of large catastrophic events.

**Acknowledgements:**

This research was sponsored by the National Natural Science Foundation of China (https://isis.nsfc.gov.cn/portal/index.asp。Grant No. 40771061). The support is gratefully acknowledged. I would like to thank my mathematician friend, Juwang Hu, for assistance in mathematical demonstration.

# Appendices

## 1. The multifractal characterization of complex Cantor set

Two sets of parameters are always employed to characterize a multifractal system. One is the *global* parameters, and the other, the *local* parameters. The global parameters include the generalized correlation dimension and the mass exponent, and the local parameters comprise the Lipschitz-Hölder exponent and the fractal dimension of the set supporting this exponent. For the two-scale Cantor set, the mass exponent ($\tau(q)$) is

$$\tau(q) = -\frac{\ln[p^q + (1-p)^q]}{\ln 3}, \quad (a1)$$

where $q$ denotes the moment order ($-\infty < q < \infty$), differing from the exponent in equation (11) in text, and $p$ is a probability measurement. Take derivative of equation (a1) with respect to $q$ yields the Lipschitz-Hölder exponent of singularity in the form

$$\alpha(q) = \frac{d\tau}{dq} = -\frac{1}{\ln 3} \frac{p^q \ln p + (1-p)^q \ln(1-p)}{p^q + (1-p)^q}. \quad (a2)$$

in which $\alpha(q)$ refers to the singularity exponent. By the Legendre transform, we have the fractal dimension of the subset supporting the exponent $\alpha(q)$ such as

$$f(\alpha) = q\alpha(q) - \tau(q)$$
$$= \frac{1}{\ln 3}[\ln[p^q + (1-p)^q] - \frac{p^q \ln p^q + (1-p)^q \ln(1-p)^q}{p^q + (1-p)^q}], \quad (a3)$$

where $f(\alpha)$ denotes the local dimension of the multifractal set. Further, the general fractal dimension spectrum ($D_q$) can be obtained in the form

$$D_q = \begin{cases} -\dfrac{p \ln p + (1-p)\ln(1-p)}{\ln 3}, & q = 1 \\ \dfrac{\tau(q)}{q-1}, & q \neq 1 \end{cases}, \quad (a4)$$

If the order moment $q \neq 1$, the general dimension can also be given by

$$D_q = \frac{1}{q-1}[q\alpha(q) - f(\alpha)]. \quad (a5)$$

If $a=0.375$ as given, then $p=a/(2/3)=0.5625$. By using the above formula, the multifractal dimension spectra and the related curves are given as follows (Figure A1, Figure B2). The capacity dimension is $D_0 \approx 0.631$, the information dimension is $D_1 \approx 0.624$, and the correlation dimension is $D_2 \approx 0.617$.

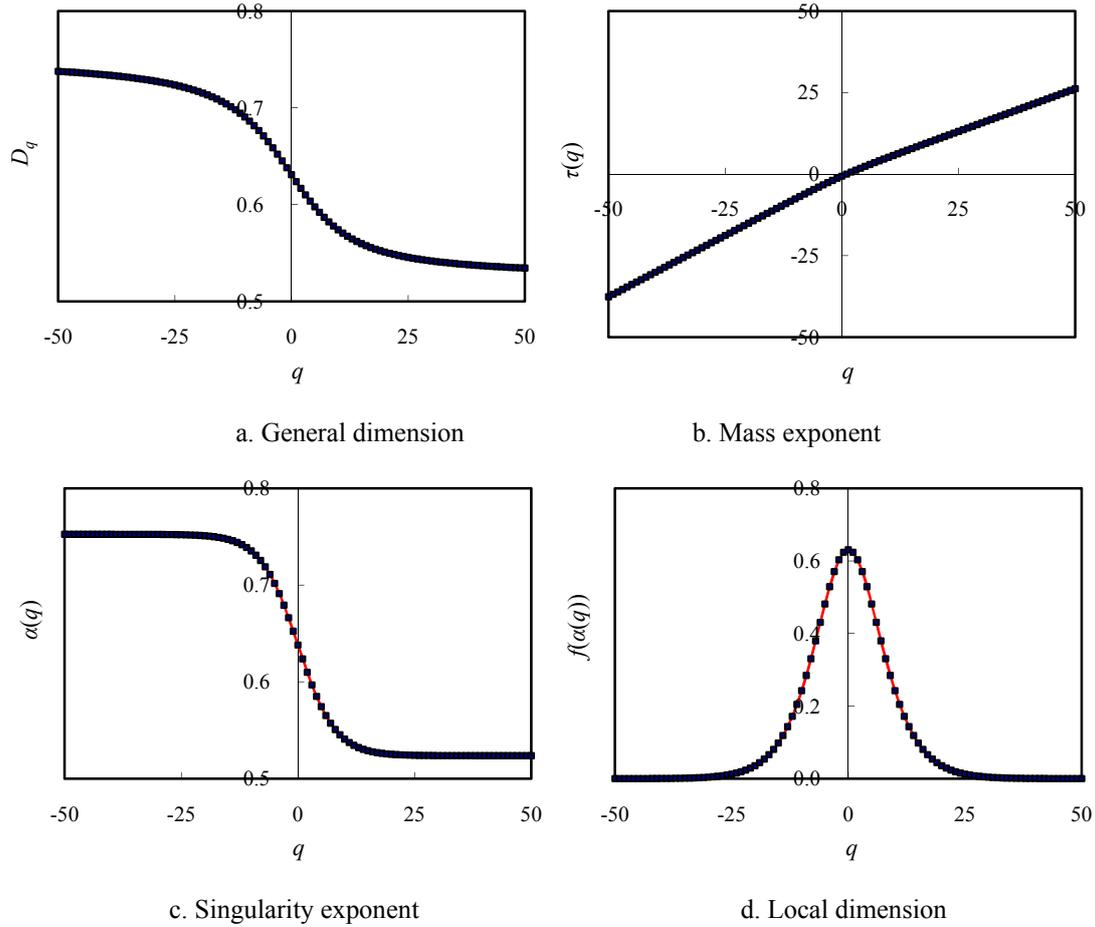

**Figure A1** The dimension spectrums of multifractal Cantor set and the curves of related parameters

($p$=0.5625)

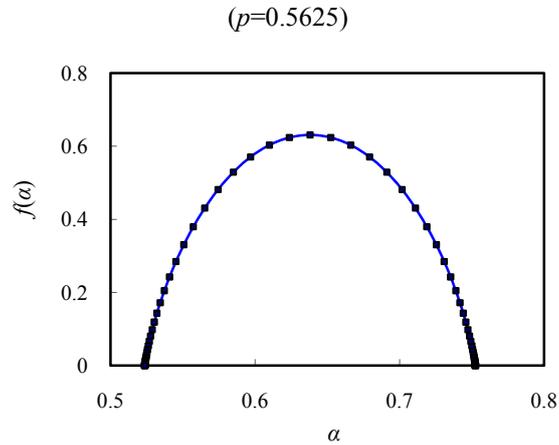

**Figure A2** The $f(α)$ curves of the local dimension v.s. the singularity exponent

## 2. A case of language hierarchy in the world

There are 15 top languages by population in the world such as Chinese, English, and Spanish. Gleich *et al* (2000) gave list of the 15 languages by number of native speakers. The population size follows the rank-size rule and the regression model is

$$P_k = 913.586 k^{-0.993},$$

where $k$ refers to rank, and $P_k$, to the population speaking the language ranked $k$, the goodness of fit is about $R^2$=0.979 (Figure B1). The fractal dimension is estimated as $D≈1.007$.

Table B1 The self-similar hierarchy of the 15 top languages by population

Unit: million

| Level | Number | Language and population | | | | Total population | Average population | Size ratio |
|---|---|---|---|---|---|---|---|---|
| 1 | 1 | Chinese | 885 | | | 885 | 885 | |
| 2 | 2 | English | 470 | Spanish | 332 | 802 | 401 | 2.207 |
| 3 | 4 | Bengali | 189 | Portuguese | 170 | 711 | 177.75 | 2.256 |
| | | Indic | 182 | Russian | 170 | | | |
| 4 | 8 | Japanese | 125 | Korean | 75 | 657 | 82.125 | 2.164 |
| | | German | 98 | French | 72 | | | |
| | | Wu-Chinese | 77 | Vietnamese | 68 | | | |
| | | Javanese | 76 | Telugu | 66 | | | |

**Source**: Gleich M, Maxeiner D, Miersch M, Nicolay F (2000). *Life Counts: Cataloguing Life on Earth*. Berlin: Springer-Verlag.

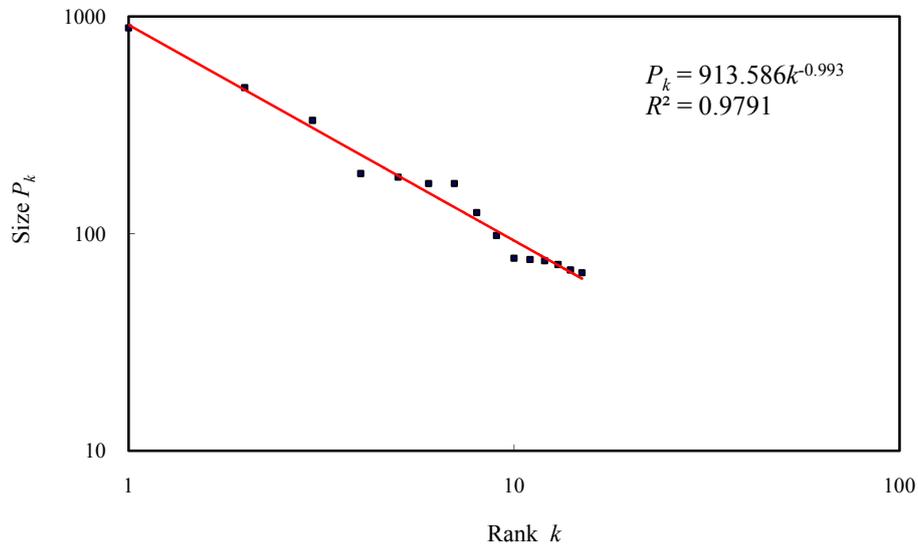

**Figure B1** The rank-size pattern of the top 15 languages by population (This is similar to a multifractal hierarchy)

According to the $2^n$ rule, the 15 languages fall into 4 classes by size (Table B1). In the top level, 1 language, i.e., Chinese, and the total of Chinese-speaking population is 885 million; in the second level, two languages, English and Spanish, with total population 802 million.... The number ratio is defined as $r_f=2$. The corresponding size ratio is around $r_p=2.209$. Thus, the fractal dimension can be estimated as $D=\ln r_f/\ln r_p=\ln(2)/\ln(2.209)=0.875$. A regression analysis yields a hierarchical scaling relation between language number, $f_m$, and average population size, $P_m$, such as

$$f_m = 371.504 P_m^{-0.872},$$

The correlation coefficient square is $R^2=0.9999$, and the fractal dimension is about $D=0.872$ (Figure B2). This is very close to 0.875, the result from size and number ratios. Because the sample is too small, the hierarchical scaling exponent is not very close to the rank-size scaling exponent, 1.007. However, if we use the lower limits of population size $a_1=520$, $a_2=260$, $a_3=130$, and $a_4=65$ to classify the languages, the corresponding number of languages is $f_1=1$, $f_2=2$, $f_3=4$, and $f_4=8$, and the scaling exponent is just 1. This suggests that if the sample is not large, we should employ more than one method to estimate the fractal parameter.

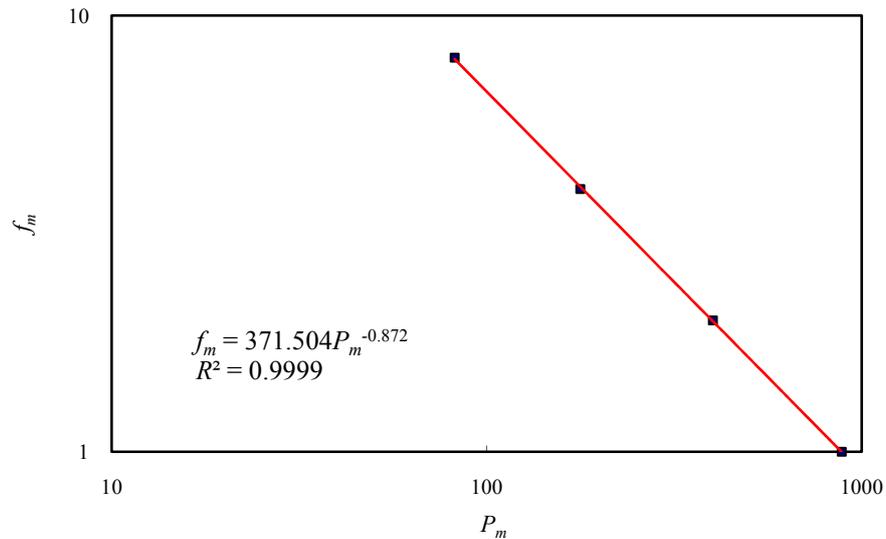

**Figure B2** The hierarchical scaling pattern of the population size speaking 15 kinds of languages